# Improved visual function in a case of ultra-low vision following ischemic encephalopathy following transcranial electrical stimulation; A case study

Ali-Mohammad Kamali[1,2,3,4], Mohammad Javad Gholamzadeh [3,4],

Seyedeh Zahra Mousavi[3,4], Maryam Vasaghi Gharamaleki [3,4], Mohammad Nami [1,2,3,5]*

[1]Department of Neuroscience, School of Advanced Medical Sciences and Technologies, Shiraz University of Medical Sciences, Shiraz, Iran
[2]DANA Brain Health Institute, Iranian Neuroscience Society-Fars Branch, Shiraz, Iran.
[3]Neuroscience Laboratory, NSL (Brain, Cognition and Behavior), Department of Neuroscience, School of Advanced Medical Sciences and Technologies, Shiraz University of Medical Sciences, Shiraz, Iran
[4]Students' Research Committee, Shiraz University of Medical Sciences, Shiraz, Iran
[5]Academy of Health, Senses Cultural Foundation, Sacramento, CA, USA

*Corresponding author*
*Mohammad Nami. Department of Neuroscience, School of Advanced Medical Sciences and Technologies, Shiraz University of Medical Sciences, Shiraz, Iran. torabinami@sums.ac.ir

## Abstract

**Objectives:** Cortical visual impairment is amongst the key pathological causes of pediatric visual abnormalities predominantly resulting from hypoxic-ischemic brain injury. Such an injury results in profound visual impairments which severely impairs patients' quality of life. Given the nature of the pathology, treatments are mostly limited to rehabilitation strategies such as transcranial electrical stimulation and visual rehabilitation therapy.
**Case description:** Here, we discussed an 11 year-old girl case with cortical visual impairment who underwent concurrent visual rehabilitation therapy and transcranial electrical stimulation resulting in her improved visual function.
**Conclusions:** This novel and noninvasive therapeutic intervention has shown potential for application in neuro-visual rehabilitation therapy (nVRT).

## Public Significance Statement

Neurocognitive and neurobehavioral approaches in assisting individuals with ultra-low vision have recently gained more clinical attention. On the other hand, the advent of neurotechnology has opened new windows of hope in neuro-visual rehabilitation therapy (nVRT). The present report highlights the clinical outcome in a case of cortical visual impairment using transcranial electrical stimulation and VRT.

## Introduction

Cortical Visual Impairment (CVI) is amongst the core pathological entities in pediatric visual abnormalities. The condition is mainly resulted from hypoxic-ischemic brain injury and as well as head trauma and epilepsy (Good, Jan, Burden, Skoczenski, & Candy, 2001). The ischemic brain injury man encompass various visual impairments such as central or peripheral visual loss, perceptual problems, and abnormalities in eye motility (Hanna, Hepworth, & Rowe, 2017; L. Hepworth et al., 2015). It can potentially impairs quality of life and limit daily functions, leading to depression (L. R. Hepworth et al., 2015).

Meanwhile, treatments are limited and most attentions are toward rehabilitation strategies. A number of neurorehabilitation methods have been suggested to remediate visual loss and improve the vision-related quality of life. Such approaches include transcranial Direct Current Stimulation (tDCS), repetitive transorbital Alternating Current Stimulation (rtACS), repetitive Transcranial Magnetic Stimulation (rTMS), and Visual Rehabilitation Therapy (VRT) (Raimund Alber, Cardoso, & Nafee, 2015; R Alber, Moser, Sabel, & Gall, 2015). Among such methods, transcranial electrical stimulation paradigms (tES) are non-invasive neuromodulatory methods the potential to alter cortical excitability through low amplitude electrical current via stimulating targeted brain regions (Heinrichs-Graham, McDermott, Mills, Coolidge, & Wilson, 2017).

Stimulating the occipital cortex using tACS or tDCS has shown to be effective in improving visual tasks performance (Castellano et al., 2017). From the clinical perspective, few previous studies have applied brain stimulation for rehabilitation of ischemic vision loss (Raimund Alber, Moser, Gall, & Sabel, 2017). However, given the paucity of empirical data, more precise investigations need to be done to endorse the efficacy of tES' in CVI.

In this report, we present an 11 year-old girl who was suffering from ultra-low vision following occipital ischemic insult and ischemic optic neuropathy (ION). The subject underwent neurovisual rehabilitation using the combination of tES and VRT.

## Case Presentation

N.O. was an 11 year-old girl suffering from ION and occipital ischemia since the age of 6 after she survived a month of intensive care unit admission following a bout of prolonged seizure attack. Her brain MRI showed bilateral symmetrical gliosis in occipital cortices and subcortical areas as well as optic radiations indicating hypoxic changes. Eye examination was performed by ophthalmologist before treatment. The examination was within normal limits except for strabismus, visual field defect, and decreased visual acuity (VA). VA measures for near sight were 4/20 and 8/20 for left and right eyes, respectively.

Before stimulation, she underwent quantitative electroencephalography (QEEG) assessment using a 19-channel QEEG Mitsar™ setup with concurrent photic stimulation (1-36 Hz, white flashes at 40 cm distance) with visio-coritcal response assessment. The Cz electrode (10-20 system) was selected as reference and EEG data was acquired through the following protocol: 1- three minutes recording in "lights out" resting state, 2- three minutes in "lights on" resting state (90 seconds eyes closed and 90 seconds eyes open) and 3- three minutes photic stimulation (red, white, red /white- light stimulation) at 1 to 36 Hz.

Background EEG activity was normal in "lights on/off" states with no epileptic activity or abnormal discharges. Other findings were fronto-central theta coherence in "lights on" state, FP1 beta-2 activity in "lights on" state, increased biooccipital high-beta, and low-gamma amplitudes at 25 Hz photic stimulation.

Based on the QEEG finding the therapy protocol was formulated by a clinical neuroscientist and a consultant ophthalmologist. The treatment protocol comprised five sessions of tDCS in the morning followed by tACS in the afternoon over five consecutive days (April, 2018). The neuromodulation comprised: 1- dual channel tDCS (Neurostim 2™, Medina Teb, Tehran) with anodes placed on left and right occipital area (O1 and O2) and cathodes placed on left and right shoulders using a 2 mA current for 25 minutes; 2- tACS (Neurostim 2™, Medina Teb, Tehran) with electrodes positioned at FP1 and FP2, and bilaterally on maxilla using a 25 Hz (the QEEG-informed oscillation which induced maximal event related desynchronization upon photic stimulation), 1.5 mA current for 25 minutes and the offset of zero. In addition, the patient was prescribed nootropics and eye supplements including sodium valproate 200mg qd, Modafinil 100mg qd, and PRESERVISION 3™ 1 capsule qd.

The post-tES QEEG indicated increased high-beta and low-gamma relative power in the left Brodmann area 17 (primary visual cortex) upon Resting State (RS), improved left temporo-occipital beta-3 coherence in RS, and increased low-gamma amplitude upon 36 Hz photic stimulation compared to the pre-stimulation recording. This suggests an uptrained gamma activity in the primary visual cortex and enhanced beta-3 coherence within the visual network. The VA was measured following every session and turned to be 12/20 and 16/20 for left and right eyes at the end of the final session (session 10), respectively.

The patient's follow-up involved the use prescribed drugs, a self-training eye exercise named Fit Eye at 40 cm from a 15 inches monitor, luminance 60% , done by dominant hand, in dark room for thirty minutes and thirty minutes later in a well-lit room, every day for two months (https://www.brainturk.com/), and mirror-tracing task which involved tracking an spiral maze in a mirror at various luminance levels and distance for 30 minutes every other day over a 60-day follow-up period.

The patient completed the questionnaire of Vision-Related Quality of Life (VRQOL) before treatment and after two months follow-up which indicating amelioration in her VRQOL.

 The drug consumption compliance was complete whereas vision exercise task compliance was nearly 50%. In the most recent follow-up session (June, 2018) her VA was 12/20 and 14/20 for left and right eyes, respectively. The observations were consistent with subjective improvement and family satisfaction.

Pre- and Post-tES current density values were analyzed using IBM SPSS statistics V.22. Wilcoxon signed rank test. The analysis revealed thatthat post-tES values (mean= 1.22, SD= 0.11) was significantly higher than those of the pre-tES (mean= 2.70, SD= 0.05) ($p$ = 0.028, CI=95%) (Figure 1A). In addition, our study showed that the current spectral density z-score (25 Hz) within the visual cortex and z-score coherence value represented by the connectivity network was increased following 10 sessions tES (Figure 1B and 1C).

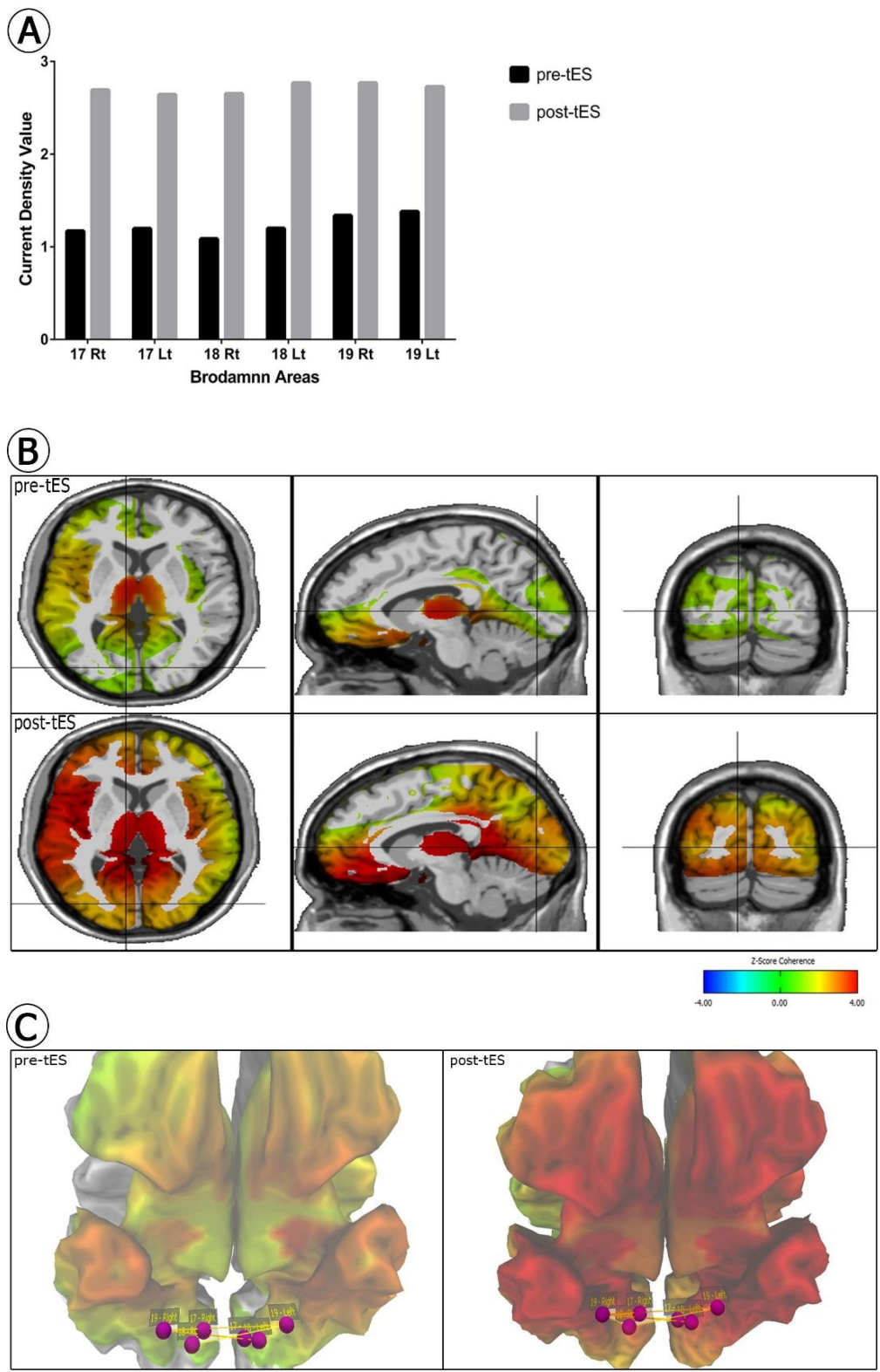

**Figure 1.** Current Spectral Density (CSD) map and functional connectivity pattern within the visual cortex. These computer-generated images using the Neuroguide Neuronavigator (Applied Neuroscience Inc. 2018) illustrate spatial distribution of the current density at 25 Hz oscillation in visual cortex prior to neuromodulation and following intervention. Panel A shows that the CSD center value at Brodmann areas 17, 18 and 19 has significantly gained in both hemispheres following a course of transcranial electrical stimulation (tES) using Wilcoxon signed rank test ($\underline{p}$ < 0.05, CI=95%) (see methods). Panel B and C demonstrate CSD z-score (25 Hz) pre- and post-tES within the visual cortex and z-score coherence value represented by the connectivity network.

*Discussion*

Here we report a case with ischemia-induced CVI who underwent concurrent tES and VRT. Though there is no definite cure for this impairment, available restoration therapies may improve visual impairments in case-by-case basis (Ospina, 2009). In addition, these patients may also suffer from related ischemic brain injuries such as ION. Similarly, there is no established treatment for ION. Previous researches have tried some surgical and medical procedures while only modest visual improvement was reported (Biousse & Newman, 2015).

Since alterations in cortical excitability and its neuroplasticity result in many neuropsychiatric diseases, tDCS and tACS might also be therapeutic options to potentially modify cortical activity in conditions such as cerebral stroke-related vision loss (Raimund Alber et al., 2015; Gall et al., 2015). Some studies show the beneficial effect of applying tDCS in addition to VRT in improving patient's visual field. In addition, changing the polarity in different frequencies in tACS can produce a sinusoidal current which can affect brain oscillations (Raimund Alber et al., 2015). The alternating current can induce phosphene in a frequency-dependent manner. That said, it might activate the visual cortex (Gall et al., 2011). Gall et al. showed that vision-related quality of life and visual field but not visual acuity improved after applying rtACS in patients with optic nerve damage (Gall et al., 2016).

Another component of our intervention was medical therapy with sodium valproate and Modafinil. Valproic acid was chosen due to its potential effects on neurogenesis and neuroprotection (Chu et al., 2015). Modafinil is an stimulant used for narcolepsy and also retains potential neuro-enhancing properties (Brandt, Ellwardt, & Storch, 2014).

In addition, we applied mirror tracing task as another rehabilitation method. Lebar et al. compared Visual-Evoked Potentials (VEPs) between the resting state and the while the patient performed mirror tracing, which involves different processes such as visuomotor transformation, hand motions perception, and attention to line orientation. The results showed that the amplitude of VEPs increases while the patients practically involve in the tasks (Lebar, Bernier, Guillaume, Mouchnino, & Blouin, 2015).

In general, application of a novel method with proper safety profile resulting in a sustained improvement in visual function (at least over a 60-day follow-up) was the highlight of our study. Most of the previous studies have articulated the effect of common neurovisual rehabilitation strategies in adults. Since cortical vision impairment is an important cause of vision loss in children, it is also necessary to evaluate and recommend the rehabilitation approaches in children and young adults (Waddington & Hodgson, 2017). Furthermore, most of the studies assess the efficacy of tDCS but not tACS and there are few studies which evaluate the therapeutic effect of concurrent tACS and tDCS in improving patient's vision. Both tACS and tDCS are relatively safe and previous researches revealed that most possible adverse effects are mild and transient. A few studies reported limited skin irritation in a small fraction of patients treated with tDCS (Matsumoto & Ugawa, 2017). Furthermore, the treatment was followed by self-training practices including mirror tracing activity and Fit Eye exercise in addition to medications to potentiate and maintain the therapeutic effect. The outcome was sustainable after two months of follow-up. Meanwhile, more investigation is needed to assess the patient's vision improvement in a long-term period.

In addition, the employment of noninvasive brain stimulation as a therapeutic approach for visual rehabilitation has not the case in our setting yet. Randomized sham-controlled studies would be required to support this initial step towards developing inexpensive, effective, and available rehabilitation protocols for resource-limited communities.

However, the use of visual acuity alone to assess the patient's vision was one of the limitations of our study and other assessments such as background and after treatment perimetry should be included in further investigations. Lack of the patient's full compliance in performing her exercises over a relatively short (260 days) follow-up was another limitation.

In conclusion, since tES is generally regarded as a safe and tolerable approach in neurorehabilitation with a shorter treatment period compared to the functional rehabilitation therapy, it might potentially be considered as alternative or add-on therapeutic approaches to remediate such patient's vision. Further studies i.e. randomized controlled clinical trials are needed not only to examine the efficacy of tES in CVI, but also to define and suggest the most appropriate neurostimulation protocols in cases with similar neurovisual profile.

### Acknowledgments
We thank the DANA Brain Health Institute; Iranian Neuroscience Society, Fars Chapter, Shiraz, Iran

### Authorship
All named authors meet the International Committee of Medical Journal Editors (ICMJE) criteria for authorship for this article. The authors take responsibility for the integrity of the work as a whole, and have given their approval for this version to be published.

### Authors' contributions
The authors declare that the research was conducted in the absence of any commercial or financial relationships that could be construed as a potential conflict of interest. All authors equally contributed to this work and they read and approved the final manuscript.

### Compliance with Ethics Guidelines
This case is reported with the informed consent of the patient.

### Data Availability
The datasets generated during and/or analyzed during the current study are available from the corresponding author on reasonable request.